\documentclass[preprint]{aastex6}
\fullcollaborationName{The Friends of AASTeX Collaboration}

\begin{document}

%% LaTeX will automatically break titles if they run longer than
%% one line. However, you may use \\ to force a line break if
%% you desire.

%\title{YONSEI EVOLUTIONARY POPULATION SYNTHESIS (YEPS).\\ II. SPECTRO-PHOTOMETRIC EVOLUTION OF HELIUM-ENHANCED STELLAR POPULATIONS}
\title{Yonsei evolutionary population synthesis (YEPS).\\ II. Spectro-photometric evolution of helium-enhanced stellar populations}

%% Use \author, \affil, plus the \and command to format author and affiliation 
%% information.  If done correctly the peer review system will be able to
%% automatically put the author and affiliation information from the manuscript
%% and save the corresponding author the trouble of entering it by hand.
%%
%% The \affil should be used to document primary affiliations and the
%% \altaffil should be used for secondary affiliations, titles, or email.

%% Authors with the same affiliation can be grouped in a single
%% \author and \affil call.
\author{Chul Chung\altaffilmark{1}, Suk-Jin Yoon\altaffilmark{1,2}, and Young-Wook Lee\altaffilmark{1,2}}
%% Notice that each of these authors has alternate affiliations, which
%% are identified by the \altaffilmark after each name.  Specify alternate
%% affiliation information with \altaffiltext, with one command per each
%% affiliation.

\altaffiltext{1}{Center for Galaxy Evolution Research, Yonsei University, Seoul 03722, Korea; chulchung@yonsei.ac.kr}
\altaffiltext{2}{Department of Astronomy, Yonsei University, Seoul 03722, Korea; sjyoon0691@yonsei.ac.kr}

%% Mark off the abstract in the ``abstract'' environment. 
\begin{abstract}

The discovery of multiple stellar populations in Milky Way globular clusters (GCs) has stimulated various follow-up studies on helium-enhanced stellar populations.
Here we present the evolutionary population synthesis models for the spectro-photometric evolution of simple stellar populations (SSPs) with varying initial helium abundance ($Y_{\rm ini}$).
We show that $Y_{\rm ini}$ brings about {dramatic} changes in spectro-photometric properties of SSPs.
Like the normal-helium SSPs, the integrated spectro-photometric evolution of helium-enhanced SSPs is also dependent on metallicity and age for a given $Y_{\rm ini}$.
{We discuss the implications and prospects for the helium-enhanced populations in relation to the second-generation populations found in the Milky Way GCs.} 
All of the models are available at \url{http://web.yonsei.ac.kr/cosmic/data/YEPS.htm}.

\end{abstract}

%% Keywords should appear after the \end{abstract} command. 
%% See the online documentation for the full list of available subject
%% keywords and the rules for their use.

%% From the front matter, we move on to the body of the paper.
%% Sections are demarcated by \section and \subsection, respectively.
%% Observe the use of the LaTeX \label
%% command after the \subsection to give a symbolic KEY to the
%% subsection for cross-referencing in a \ref command.
%% You can use LaTeX's \ref and \label commands to keep track of
%% cross-references to sections, equations, tables, and figures.
%% That way, if you change the order of any elements, LaTeX will
%% automatically renumber them.

%% We recommend that authors also use the natbib \citep
%% and \citet commands to identify citations.  The citations are
%% tied to the reference list via symbolic KEYs. The KEY corresponds
%% to the KEY in the \bibitem in the reference list below. 

\section{Introduction} \label{sec:intro}

The study of helium-enhanced stellar populations in the Milky Way globular clusters (GCs) was initiated by the discovery of multiple red giant branches (RGBs) and  multiple main-sequences (MSs) in $\omega$~Cen \citep[e.g.,][]{1999Natur.402...55L, Piot02, 2004ApJ...605L.125B}.
Using helium-enhanced stellar populations, \citet{2004ApJ...612L..25N} reproduced the multiple MSs in ${\omega}$~Cen, and \citet{Lee05b} reproduced the multiple MSs and extremely blue horizontal-branch (HB) stars simultaneously in ${\omega}$~Cen and NGC~2808.
These results stimulated the study of multiple stellar populations in the Milky Way GCs and led the explosive increase of subsequent studies \citep[e.g.,][]{2005ApJ...621..777P, 2007ApJ...661L..53P, 2008MNRAS.390..693D, Yoon08, 2009ApJ...697L..58A, Han09, 2013ApJ...762...36J, 2013ApJ...767..120M, 2013ApJ...775...15P, 2014ApJ...784...32K, 2015ApJS..216...19L, 2016MNRAS.457.4525T}.
The application of the helium-enhanced stellar populations to the integrated properties \citep{2011ApJ...740L..45C, 2013ApJ...769L...3C} makes it possible to understand the FUV-strong GCs in M87 \citep{2006AJ....131..866S, 2007MNRAS.377..987K, 2017MNRAS.464..713P} and the UV flux upturn phenomenon of massive early-type galaxies \citep[e.g.,][]{1988ApJ...328..440B, Park97, 1998ApJ...492..480Y, 1999ApJ...513..128Y, 1999ARA&A..37..603O, 2011ApJS..195...22Y}.
In this paper, we present the integrated colors and absorption-line indices predicted from the YEPS model \citep{2013ApJS..204....3C}, taking into account helium-enhanced subpopulations with various {initial helium} abundances.

The paper is organized as follows.
Section 2 describes how our models are constructed.
In Section 3, we present our model results of helium-enhanced SSP models for broadband colors and LICK absorption indices.
Section 4 discusses the implications of helium-enhanced stellar populations for the integrated properties of various stellar systems.

\section{Construction of models} \label{sec:const}

The models for helium-enhanced stellar populations are constructed based on the $Y^2$ stellar libraries with enhanced {initial helium} abundances \citep{2015HiA....16..247L}.
All of the other model ingredients for helium-enhanced stellar populations are the same as those for the normal-helium models described in  \citet{2013ApJS..204....3C, 2013ApJ...769L...3C}.
{As discussed in the literature \citep [e.g.,][]{2007A&A...468..657G, 2011A&A...534A...9S, 2013MNRAS.430..459D}, we assume that the enhanced helium does not affect the stellar spectra used in our models.}
{We note that our models are calculated at fixed $Z$. 
In this case, if $Y_{\rm ini}$ increases, the ${\rm [Fe/H]}$ of our models changes accordingly, keeping $Z$ fixed. 
Equations~\ref{eq1}-\ref{eq3} show how we derive ${\rm [Fe/H]}$ at given $Y_{\rm ini}$ and $Z$: 
\begin{equation}
Y=\Delta Y/\Delta Z \times Z + Y_{\rm ini},
\label{eq1}
\end{equation}
\begin{equation}
X=1.0-Y-Z,
\label{eq2}
\end{equation}
\begin{equation}
{\rm [Fe/H]}=\log_{10} \left({{Z \over X}\over{Z_{\odot} \over X_{\odot}}}\right)-0.217,
\label{eq3}
\end{equation}
where $Z_{\odot}/X_{\odot}$ is 0.025.
The value $-0.217$ is the ${\rm [Fe/H]}$ decrement when ${\rm [\alpha/Fe]}=0.3$, under the $\alpha$-elements mixture of \citet{Kim02}.}
The {initial helium} abundances of the $Y^2$ stellar libraries consist of five values of $Y_{\rm ini}=0.23$, 0.28, 0.33, 0.38, and 0.43.
We set the initial helium of the normal-helium stellar populations as $Y_{\rm ini}=0.23$ and assume the same Galactic helium enrichment parameter, $\Delta Y/\Delta Z$, to be 2.0.
Hence, the helium abundance of a $Z=0.02$ star with an {initial helium} of $Y_{\rm ini}=0.43$ would be $Y=2.0 \times 0.02 + 0.43 = 0.47$.
Table~\ref{tab:table1} summarizes the input parameters adopted in this paper{, and Table~\ref{tab:table2} provides the ${\rm [Fe/H]}$ to $Z$ conversion for $\Delta Y/\Delta Z = 2.0$ based on Equations~\ref{eq1}-\ref{eq3}}.

\begin{figure}
\includegraphics[angle=0,scale=0.65]{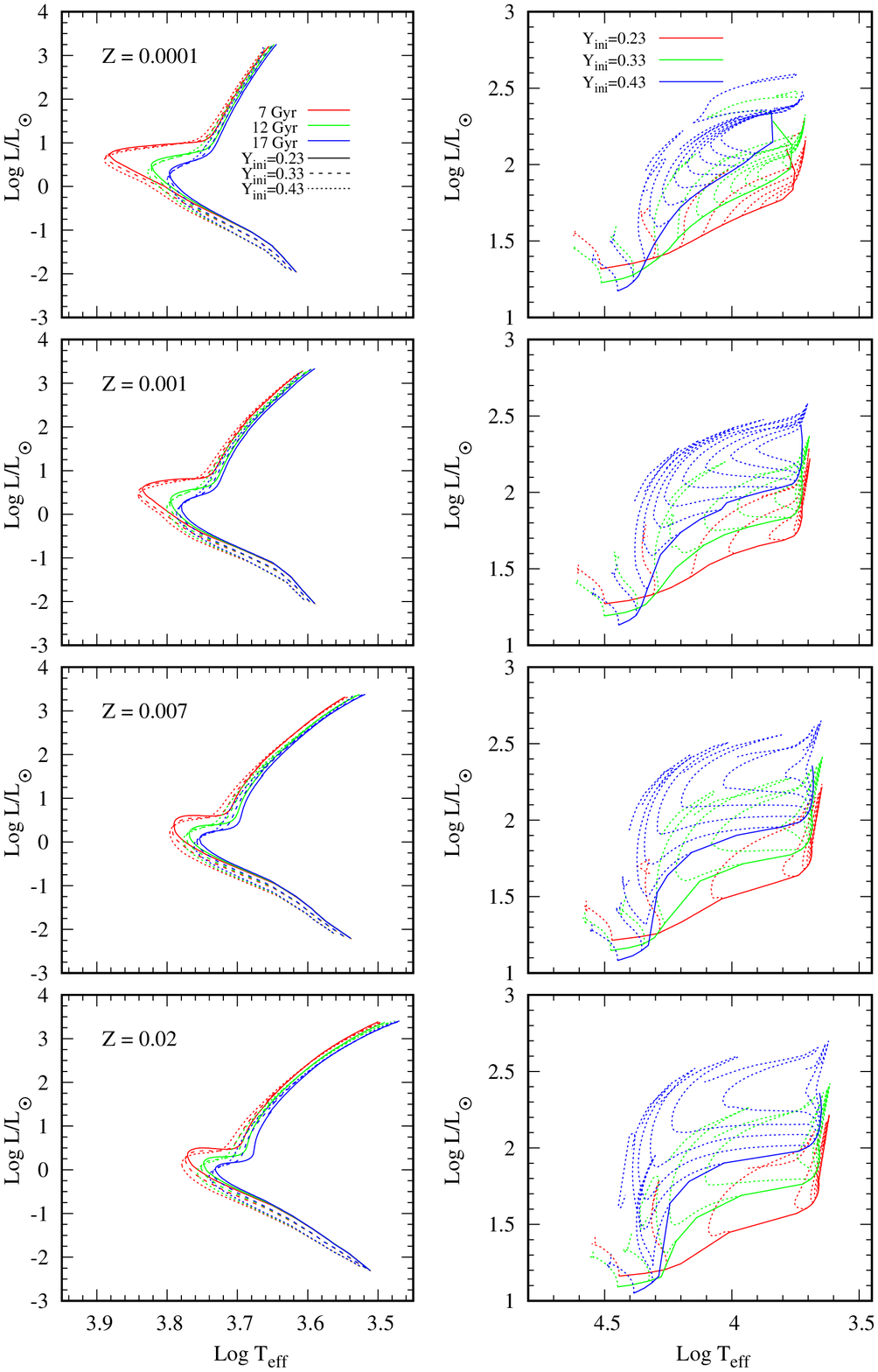}
\caption
{Effect of {initial helium} abundance on the stellar evolution.
The left panels are isochrones {with ages of 7, 12, and 17 Gyr} for metallicities from $Z$=0.0001 to $Z$=0.02 {(from top to bottom)}.
The solid lines are isochrones for normal-helium abundance ($Y_{\rm ini}=0.23$), and the {dashed and dotted} lines correspond to helium-enhanced isochrones for $Y_{\rm ini}=0.33$ and 0.43, respectively.
The right panels are the evolutionary tracks for helium burning stars with various mass.
The metallicities of the right panels are the same as those of the left panels.
The red, green, and blue lines indicate $Y_{\rm ini}=0.23$, 0.33, and 0.43, respectively.
{The solid lines are loci for zero-age HBs at given {initial helium} abundances.}
}
\label{fig1}
\end{figure}

\begin{figure}
\includegraphics[angle=0,scale=0.9]{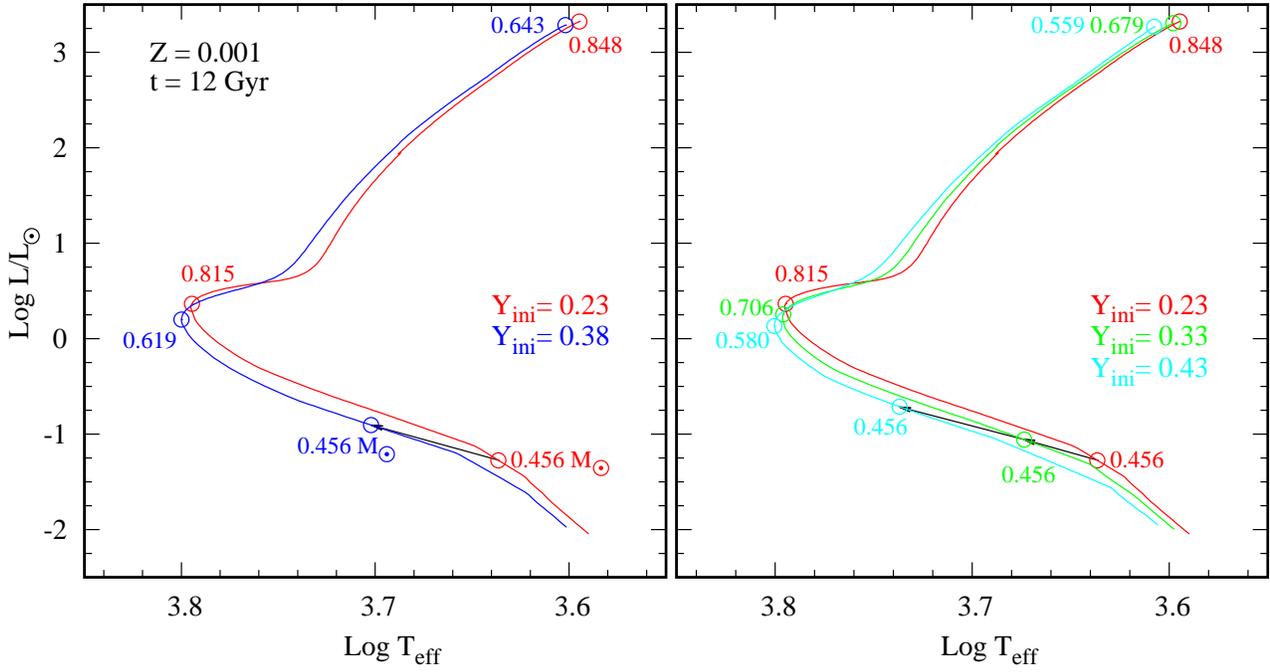}
\caption
{Effect of {initial helium} abundance on the stellar evolution from the low main-sequence to the tip of the RGB.
Left panel: the isochrones with an initial helium of $Y_{\rm ini}=0.23$ and $0.38$ for metallicities of $Z$=0.001.
The masses of the MS turn-off, the tip of RGBs, and 0.456 M$_{\odot}$ are indicated on the isochrones.
Right panel: the same as the left panel but for an initial helium of $Y_{\rm ini}=0.33$ and $0.43$.
Helium-enhanced stars become hotter and brighter at a given mass but show slightly fainter luminosity at the same evolutionary stages.
}
\label{fig2}
\end{figure}

Figure~\ref{fig1} displays the effect of helium abundance on the evolution of stars from the MS to the HB.
Since, for a given mass the core temperature of helium-enhanced stars is hotter than that of normal-helium stars, helium-enhanced stars evolve faster than normal-helium stars.
{Figure~\ref{fig2} shows the detailed version of the $Z=0.001$ case in Figure~\ref{fig1}. 
The helium-enhanced stars of a given mass are hotter and brighter on the MS than the stars with lower initial helium. 
However, since helium-enhanced stars evolve much faster than normal-helium stars, helium-enhanced stars are fainter at the same evolutionary stages such as the turn-off and the tip of RGB. 
Yet, their temperature is still hotter than that of the normal-helium stars.}
Consequently, helium-enhanced stars have a smaller mass at a given age and thus helium-enhanced stars in the MS-to-RGB stage are slightly fainter than normal-helium stars (left panels of Figure~\ref{fig1}).

The evolution of helium-enhanced stars in the HB stage is a bit different from those in the MS and RGB stages.
As shown in the right panels of Figure~\ref{fig1}, the zero-age HB locus of helium-enhanced stars are brighter than those of normal-helium stars.
However, the luminosity of the zero-age HB loci of helium-enhanced stars becomes extremely faint when the temperature of the stars reaches approximately 19,000~K ($\log T_{\rm eff} \sim 4.3$).
This is because helium-enhanced stars with such temperatures have extremely thin hydrogen-burning shells {with negligible energy output. 
Moreover, helium-rich stars have a smaller core mass, which is the major source of total energy output of these hot HB stars. 
As a result, extremely hot HB stars from helium-enhanced populations are fainter than those from normal-helium populations} \citep{Lee05b}.

\begin{figure*}
\includegraphics[angle=-90,scale=0.75]{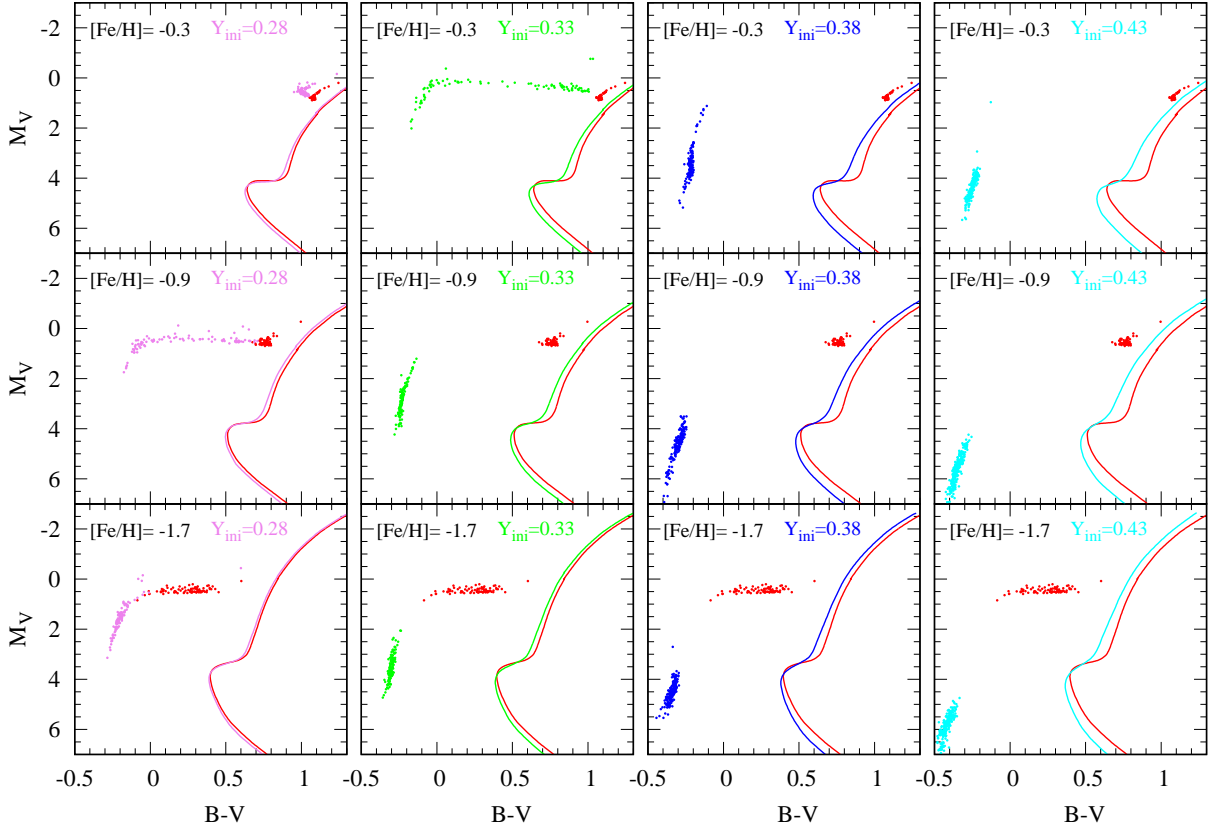}
\caption
{
Effect of {initial helium} abundance on color--magnitude diagrams (CMDs) of SSPs.
The four columns show the comparison of CMDs between SSPs with normal-helium and helium-enhanced populations.
The red, pink, green, blue, and cyan colors correspond to stellar populations for $Y_{\rm ini}=0.23$, 0.28, 0.33, 0.38, and 0.43, respectively.
The metallicities of the SSPs are $[{\rm Fe/H}]=-0.3$, --0.9 , and --1.7 from the top to the bottom of the panels.
}
\label{fig3}
\end{figure*}

The fast evolution of helium-enhanced stars exerts strong effects on the HB stage because the HB type (i.e., the mean color of HB stars) at a given age is controlled by the mean envelope mass of HB stars.
Figure~\ref{fig3} shows synthetic color--magnitude diagrams with different {initial helium} abundances and metallicities.
The adopted mass-loss efficiency parameter $\eta$ {\citep{1977A&A....57..395R}} is 0.5 and the assumed age of stellar populations is 12~Gyr.
The difference in color between normal and helium-enhanced populations on the MS and RGB stages gradually increases as the {initial helium} abundance increases.
As the {initial helium} abundance increases, the mean HB temperature increases for a given metallicity.
As a result, helium-enhanced stellar populations (i.e., $Y_{\rm ini}=0.28$, 0.33, 0.38, and 0.43) yield blue or extremely blue HB stars compared to those associated with the normal-helium stellar populations ($Y_{\rm ini}=0.23$).
Note that helium-enhanced populations show strong dependencies on the metallicity, as was shown by \citet{2013ApJS..204....3C} in normal-helium stellar populations.

\begin{figure*}
\includegraphics[angle=-90,scale=0.6]{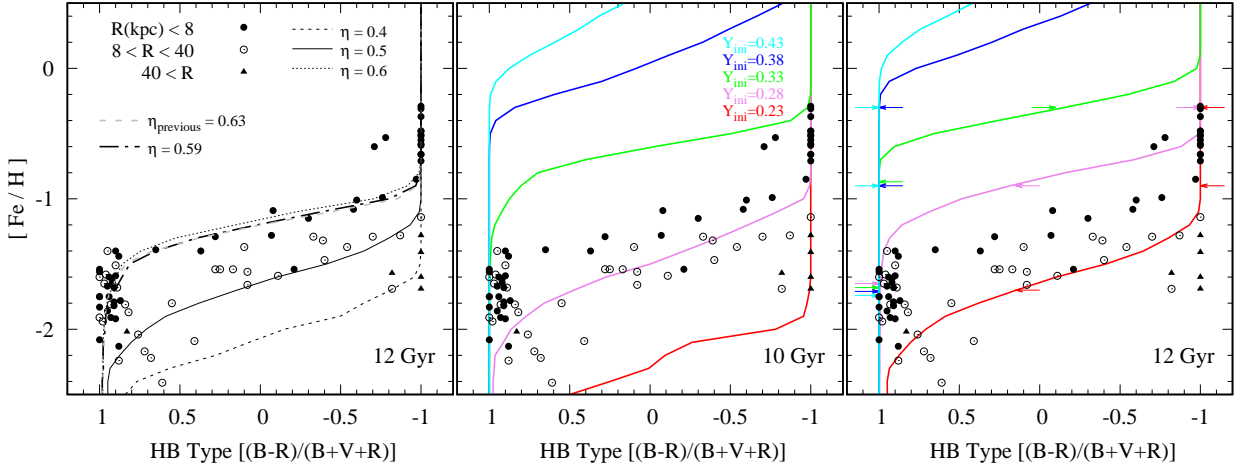}
\caption{
The mass-loss efficiency parameter ($\eta$) calibration and the effect of {initial helium} abundance on the HB morphologies of SSPs.
The left panel compares our $\eta$ calibration with that of the previous one, and at the same time shows the effect of $\eta$.
The filled circles are the Milky Way GCs within 8~kpc.
The open circles and filled triangles are GCs in the outer part of the Milky Way (larger than 8~Kpc).
The plots in the middle and right panels are the same as those in Figure~1 of \citet{2013ApJS..204....3C}  but for different {initial helium} contents with $Y_{\rm ini}=0.23$, 0.28, 0.33, 0.38, and 0.43.
The red, pink, green, blue, and cyan lines correspond to the model HB types of $Y_{\rm ini}=0.23$, 0.28, 0.33, 0.38, and 0.43, respectively.
The arrows in the right panel indicate the points that corresponds to the CMDs in Figure~\ref{fig3}.
As the {initial helium} abundance increases, the HB types of those populations show a bluer type at a given age and metallicity.
}
\label{fig4}
\end{figure*}

Figure~\ref{fig4} presents the mass-loss efficiency parameter ($\eta$) calibration \citep{1977A&A....57..395R, LDZ94} of the HB morphology to the inner-halo GCs of the Milky Way and the resulting HB types at a given age and metallicity with respect to various {initial helium} abundances.
Our previous paper \citep{2013ApJS..204....3C} in this series adopted 12~Gyr as the age of GCs in the inner halo of the Milky Way, and required $\eta$ to be 0.63.
We did this calibration using normal-helium ($Y_{\rm ini}=0.23$) stellar populations and apply the same $\eta$ to all other models of different {initial helium} abundances.
After the release of our previous models based on $Y^2$-isochrones of \citet{Kim02}, we have improved our mass-loss table following the updated version of $Y^2$-isochrones, and under this condition, $\eta$ turns out to be 0.59 for 12~Gyr.
The left panel of Figure~\ref{fig4} compares our previous $\eta$ calibration to the new one, and the two different calibrations reproduce the same HB type for the same age and metallicity.
If we assume the age of the inner halo as 13~Gyr \citep{2010ApJ...708..698D}, the new $\eta$ becomes 0.53\footnote{
The $\eta$-calibration of our model is dependent on age and $Y_{\rm ini}$.
As age and $Y_{\rm ini}$ increase at a given metallicity, the $\eta$ value that matches the HB morphology of the inner-halo GCs decreases due to the smaller mean mass of HB stars.
Therefore, if we adopt the slightly greater primodial helium abundance, the value of $\eta$ becomes smaller than that of $Y_{\rm ini}=0.23$. 
In addition, if we include helium spread in inner-halo Milky Way GCs, the increase of the average helium abundance will reduce the $\eta$ value.
}.
Throughout this paper we adopt $\eta=0.5$ as a standard and also provide models for $\eta=0.4$ and 0.6 to show the effect of the mass-loss efficiency.
In general, the effect of helium enhancement is similar to that of an age increase because the mean stellar mass also becomes smaller with older ages.
The {initial helium} abundance determines specific values of metallicity for which the transition from the blue HB type to the red HB type takes place for given ages.
For a comparison to CMDs, we place arrows of the same colors as those in Figure~\ref{fig3} in the rightmost panel of Figure~\ref{fig4}.
As {initial helium} abundance increases, our models show the parallel displacement toward the metal-rich regime in the HB type versus the metallicity plane.

It should be noted that extremely helium-enhanced stellar populations (e.g., $Y_{\rm ini} \geq 0.38$), when old ($t \geq 12$~Gyr) and metal-poor (${\rm [Fe/H]} \leq -2.0$), often have HB masses that are smaller than the core masses ($\leq 0.45 M_{\odot}$) given by the HB tracks.  
Table~\ref{tab:table3} provides the valid range of ages of helium-enhanced models that are free from such cases.

\section{Results} \label{sec:results}

\begin{figure*}
\includegraphics[angle=-90,scale=0.70]{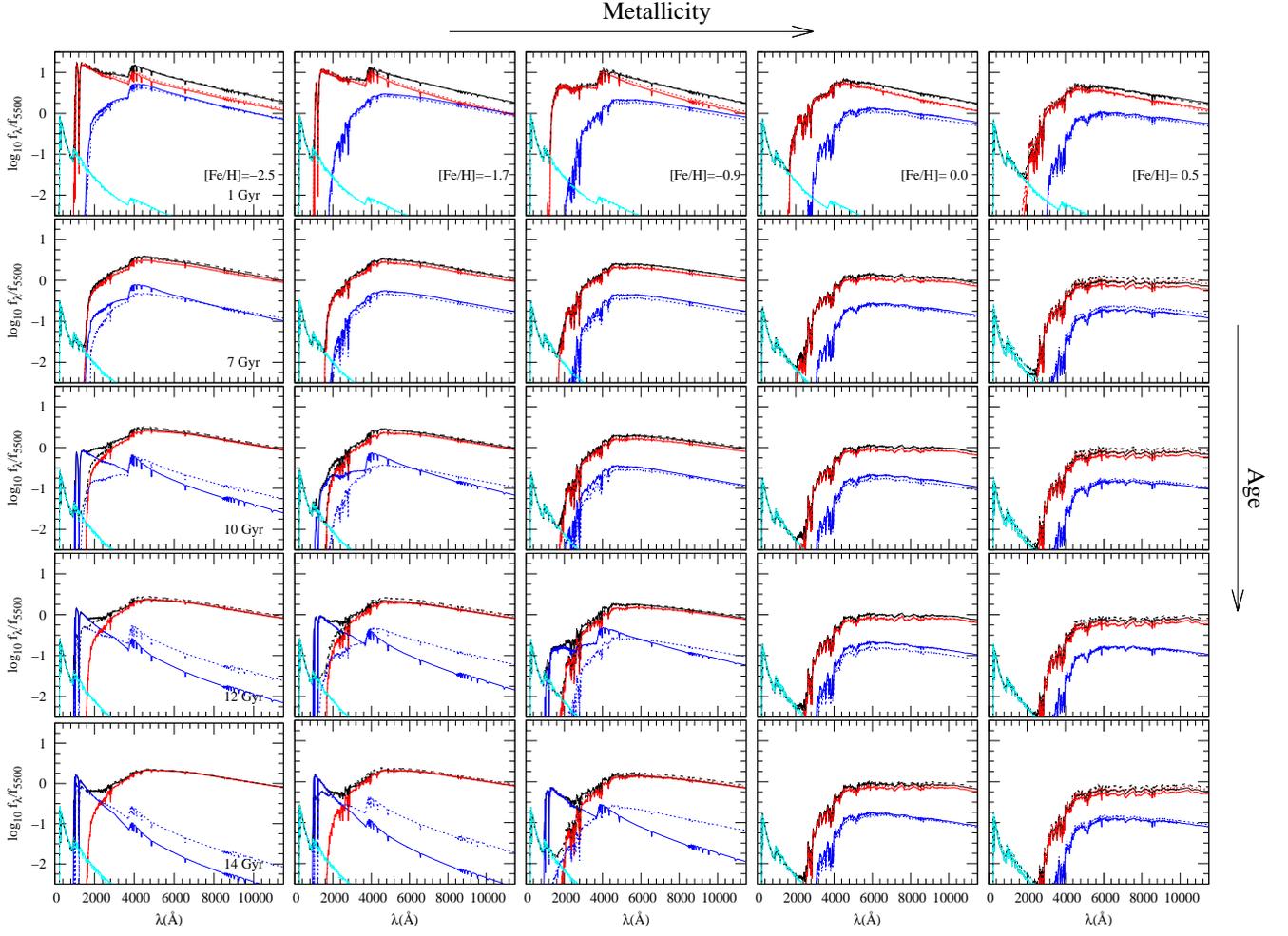}
\caption
{
Spectral energy distributions (SEDs) of simple stellar populations with $Y_{\rm ini}=0.28$.
The lines in red, blue, and cyan are the flux contributions from MS stars to RGB stars, HB stars, and post-asymptotic giant branch stars, respectively.
The dotted lines in each plot are SEDs for normal-helium abundance ($Y_{\rm ini}=0.23$) at the same conditions.
The $[\alpha/{\rm Fe}]$ of helium-enhanced stellar population in the YEPS model is 0.3 and all flux values are normalized by the flux value at 5,500~{\AA} for 12~Gyr and the $[{\rm Fe}/{\rm H}]=0.0$ model.
The ages of simple stellar populations are 1, 7, 10, 12, and 14~Gyr from the top row to the bottom row.
The metallicities are $[{\rm Fe}/{\rm H}]=-2.5$ to 0.5, as labeled from the left column to the right column.
As age increases and metallicity decreases, the effect of hot HB stars becomes dominant in the short wavelength regimes ($\leq$ 4000{\AA}).
}
\label{fig5}
\end{figure*}

\begin{figure*}
\includegraphics[angle=-90,scale=0.70]{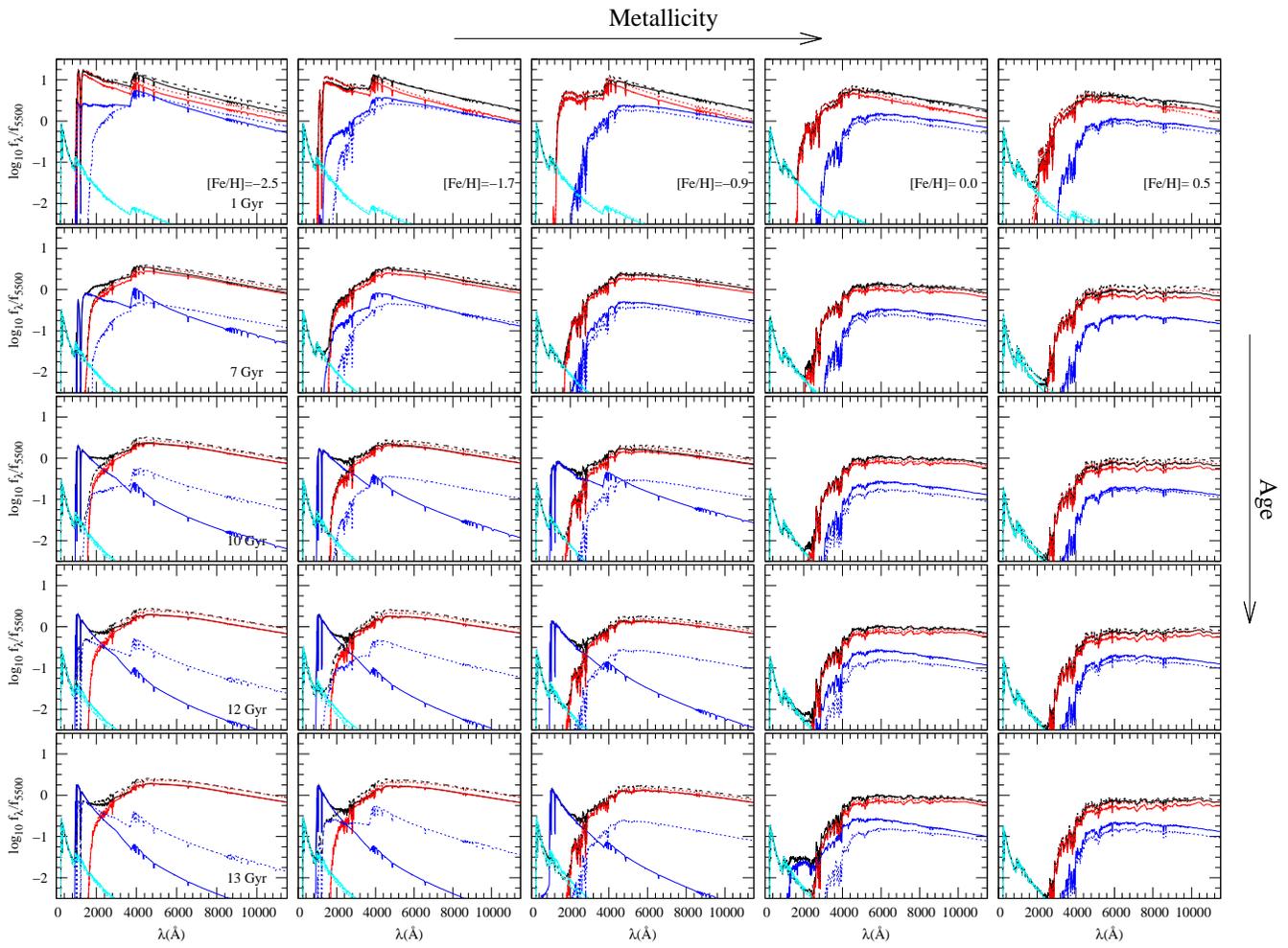}
\caption
{
Same as Figure~\ref{fig5} but for $Y_{\rm ini}=0.33$ {and ages of 1, 7, 10, 12, and 13~Gyr}.
}
\label{fig6}
\end{figure*}

\begin{figure*}
\includegraphics[angle=-90,scale=0.70]{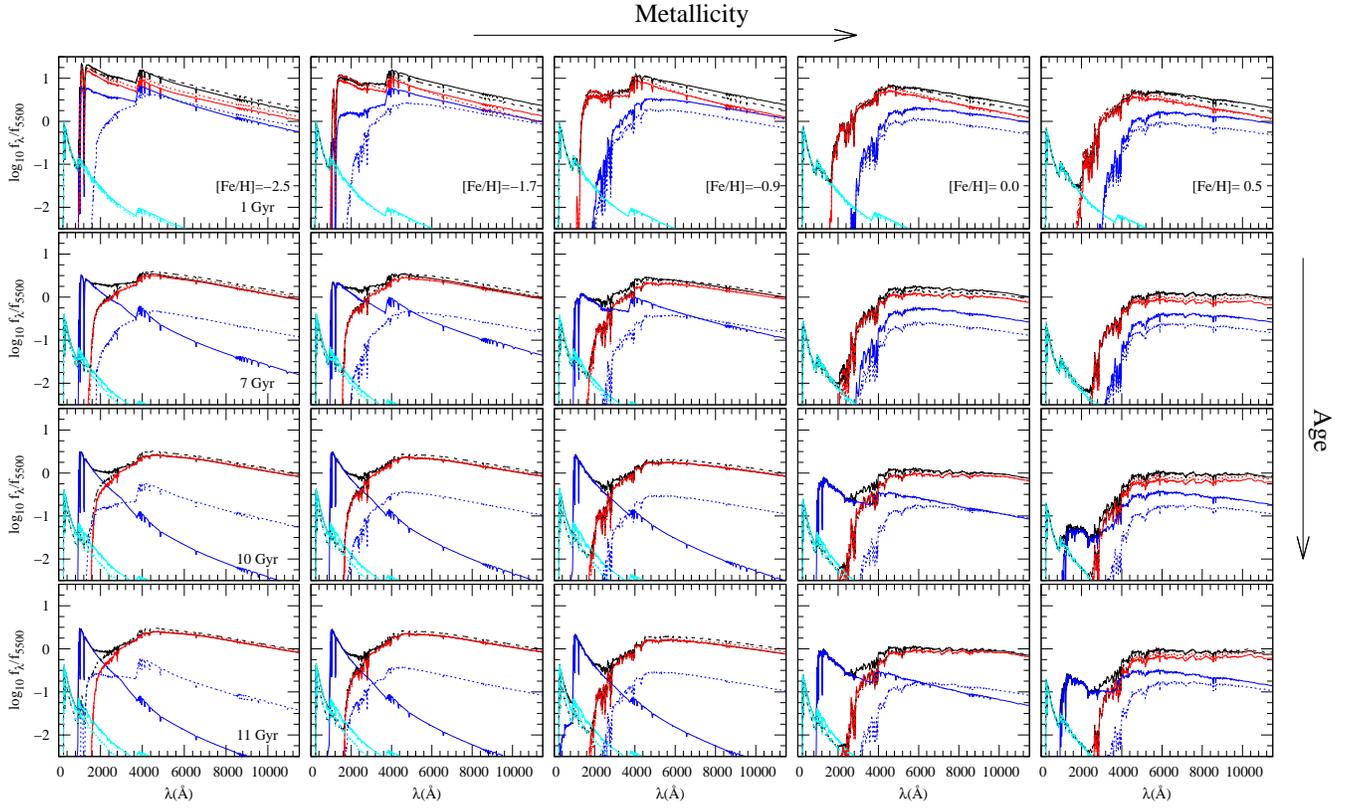}
\caption
{
Same as Figure~\ref{fig5} but for $Y_{\rm ini}=0.38$ {and ages of 1, 7, 10, and 11~Gyr}.
}
\label{fig7}
\end{figure*}

\begin{figure*}
\includegraphics[angle=-90,scale=0.70]{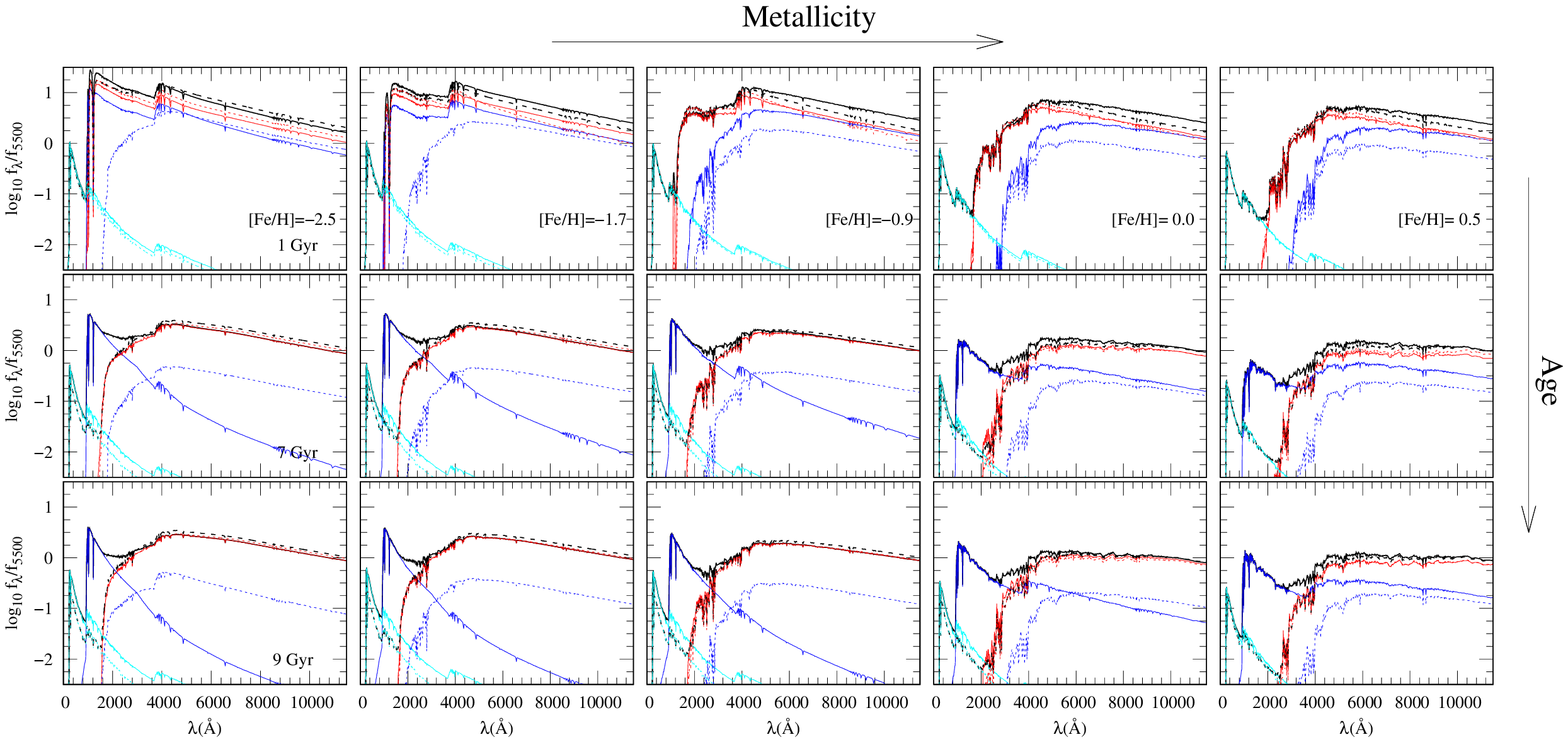}
\caption
{
Same as Figure~\ref{fig5} but for $Y_{\rm ini}=0.43$ {and ages of 1, 7, and 9~Gyr}.
}
 \label{fig8}
\end{figure*}

In Figures~\ref{fig5}-\ref{fig8}, we present integrated SEDs of the model for helium-enhanced stellar populations under different assumptions of {initial helium} abundance.
The dotted lines are SEDs for normal-helium stellar populations ($Y_{\rm ini}=0.23$), and the colors denote the fluxes from the same evolutionary stages.
It is evident from the figures that the SEDs with helium-enhanced stellar populations generally show stronger UV fluxes as age increases and metallicity decreases.
This trend is the same as that for SEDs of normal-helium stellar populations, but the absolute UV fluxes of helium-enhanced stellar populations are far greater than those of normal-helium populations.
As explained in the previous section, these strong UV fluxes come from extreme HB stars associated with helium-enhanced stellar populations.
{The ${\rm FUV}$ flux of a helium-enhanced stellar population ($Y_{\rm ini}=0.43$) with 9~Gyr and $[{\rm Fe/H}]=0.5$ is approximately two times stronger than that of normal-helium stellar populations ($Y_{\rm ini}=0.23$) with 12~Gyr and $[{\rm Fe/H}]=-2.5$ (see Figure~\ref{fig8}).}
For a normal-helium case, strong ${\rm FUV}$ flux is created only by hot, blue HB stars from low metallicity and old age stellar populations.
Thus, without extreme HB stars associated with helium-enhanced populations, there seems to be no way to generate such strong ${\rm FUV}$ flux with the metallicity as high as $[{\rm Fe/H}]=0.5$ and the young age of 9~Gyr.
On the other hand, the integrated SEDs of the MS-to-RGB stage (red lines) are not significantly different from each other.
This is because, compared to the mean temperature increase in the HB stage, the temperature difference is quite small between normal-helium and helium-enhanced stars in the MS-to-RGB stage.

Figures~\ref{fig9} and \ref{fig10} present the photometric evolution models of helium-enhanced stellar populations.
Although the extent of the variation in integrated colors is different from that of the {initial helium} abundances used in the model, the colors of helium-enhanced stellar populations become bluer as the helium abundance increases at a given metallicity \citep{2013ApJ...769L...3C}.
As expected from the SEDs in Figures~\ref{fig5}-\ref{fig8}, the most sensitive color to the helium-enhanced population is $({\rm FUV}-V)_0$.
The HB types of these stellar populations directly affect the integrated $({\rm FUV}-V)_0$ models (see Figure~\ref{fig4}).
At the age of 8~Gyr, the model of $Y_{\rm ini}=0.43$ already shows extremely blue $({\rm FUV}-V)_0$ color regardless of the metallicities.
With the combination of {initial helium} abundance and age, our models are able to give an explanation for the observed $({\rm FUV}-V)_0$ color range of GCs in M87 as blue as 3 to 8 in the AB magnitude \citep{2006AJ....131..866S}.
We refer the reader to Figure~2 of \citet{2011ApJ...740L..45C} for a detailed comparison of our model with the M87 GC observation.

Interestingly, the models with helium-enhanced stellar populations ($Y_{\rm ini}=0.28$ and 0.33) show slightly redder optical colors in the metal-poor regime at 12~Gyr, although the HB types of these populations are ``blue'' for most extreme HB stars.
This is because the luminosities of the extreme HB stars are so faint ($\Delta M_V = \left< M_{EHB} \right> - \left< M_{HB} \right> \geq 4$) that their effect on luminosity-weighted colors is even smaller than that of the usual blue HB stars in normal-helium stellar populations.
The flux contribution of hot HB stars to the infrared regime of integrated SEDs is so small that the NIR-related colors result in negligible deviation from normal-helium populations.
For instance, the effect of helium-enhanced populations becomes negligible in the color of $(V-K)_0$ when the ages are old ($t>10$~Gyr).
This means that, using $(V-K)_0$, which is insensitive to the presence of helium-enhanced stellar populations, the zero-point calibration of stellar population synthesis models for SSPs with respect to observations can be done regardless of the initial helium abundances of stellar populations.
Note that the red clump stars from the metal-rich (${\rm [Fe/H]} \geq 0.0$) and mildly helium-enhanced ($Y_{\rm ini} \leq 0.33$) populations are slightly brighter than the normal-helium red clumps \citep{2015MNRAS.453.3906L}, but are insignificant in the integrated color of $(V-K)_0$ and other colors.  
The slightly brighter luminosity ($\Delta M_K \sim 0.5$) of helium-enhanced red clumps is insufficient to result in color differences in all wavelength regimes.

Figures~\ref{fig11}-\ref{fig12} present the LICK/IDS absorption indices from the helium-enhanced stellar population models.
The effects of the helium-enhanced populations on the integrated absorption indices look similar to those on the integrated colors.
The most temperature sensitive Balmer indices (H$\beta$, H$\gamma$, and H$\delta_F$) and the indices located in the shorter wavelengths are strongly affected by hot HB stars from the helium-enhanced populations.
Particularly, Balmer indices are directly tracing the HB morphologies of the helium-enhanced populations \citep{2013ApJS..204....3C}.
The absorption strengths of Balmer indices get weaker as the age increases and the metallicity decreases.
This is because the strengths of H$\beta$ and H$\delta_F$ of a star reach peaks at  $T_{\rm eff} \sim 9500$~K.
Thus, the effect of bluer ($T_{\rm eff} > 9500$~K) or extremely blue HB stars ($T_{\rm eff} > 20,000$~K) on H$\beta$, H$\gamma$, and H$\delta$ is relatively smaller than ordinary blue HB stars whose mean temperature is around 9000~K (see Figure~\ref{fig1}).
As a result, hot and fainter HB stars lead to decreasing Balmer lines {\citep[e.g.,][]{2000AJ....120..998L, 2004ApJ...608L..33S, 2011MNRAS.412.2445P}}.

The Fe5015, Mg$_2$, Mg$b$, $\left<{\rm Fe}\right>$, and NaD absorption strengths of the helium-enhanced populations display relatively small deviation from those of the normal-helium populations compared to the Balmer indices.
These indices are merely sensitive to the amount of the elements (i.e., Fe, Mg, and Na) rather than the temperature.
As described in \citet{2013ApJS..204....3C}, warm HB stars ($T_{\rm eff}\sim 9000$~K) from the helium-enhanced stellar populations tend to depress the strength of the metal indices above.

\section{Discussion} \label{sec:discussion}

As part of the Yonsei Evolutionary Population Synthesis (YEPS) series, we have presented the models with helium-enhanced stellar populations.
The construction of the helium-enhanced models was motivated by the discovery of the multiple stellar populations in Milky Way GCs.
The helium-enhanced populations generally show bluer colors and stronger indices than the normal-helium populations in the temperature sensitive colors and absorption indices.

\begin{figure*}
\includegraphics[angle=0,scale=0.75]{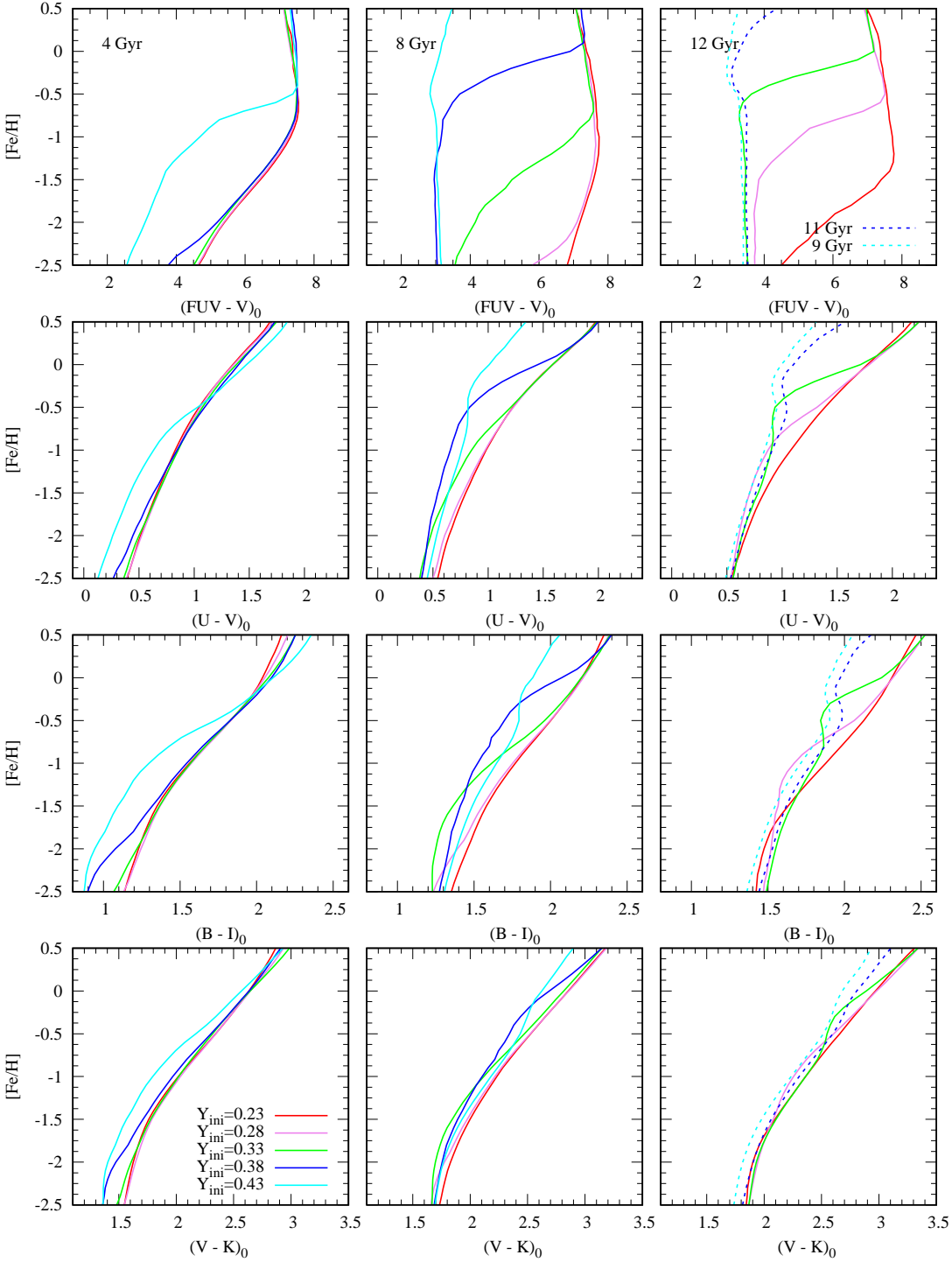}
\caption{Effect of {initial helium} on integrated colors of SSPs.
The red, pink, green, blue, and cyan lines correspond to SSPs with helium abundances of $Y_{\rm ini}=0.23$, 0.28, 0.33, 0.38, and 0.43, respectively.
The ages of SSPs are 4, 8, and 12 Gyr from the left panels to the right.
The integrated colors for $({\rm FUV}-V)_0$, $(U-V)_0$, $(B-I)_0$, and $(V-K)_0$ are presented from the top panels to the bottom.
{For $Y_{\rm ini}=0.38$ and 0.43 models, we present 11 and 9~Gyr models, respectively, in the panels of 12~Gyr.}
}
\label{fig9}
\end{figure*}

\begin{figure*}
\includegraphics[angle=0,scale=0.75]{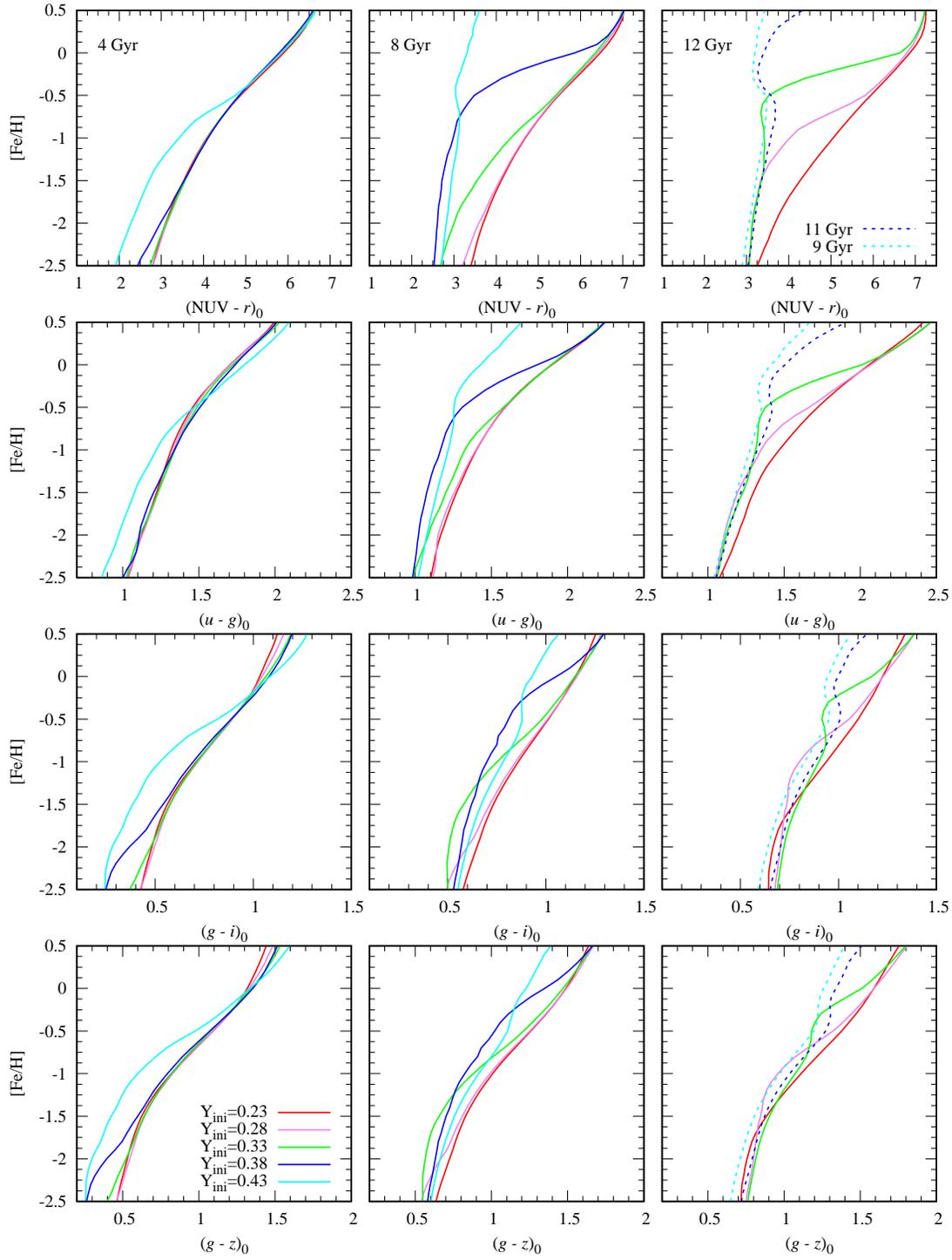}
\caption{
Same as Figure~\ref{fig9} but for colors of $({\rm NUV}-r)_0$, $(u-g)_0$, $(g-i)_0$, and $(g-z)_0$.
}
\label{fig10}
\end{figure*}

\begin{figure*}
\includegraphics[angle=0,scale=0.75]{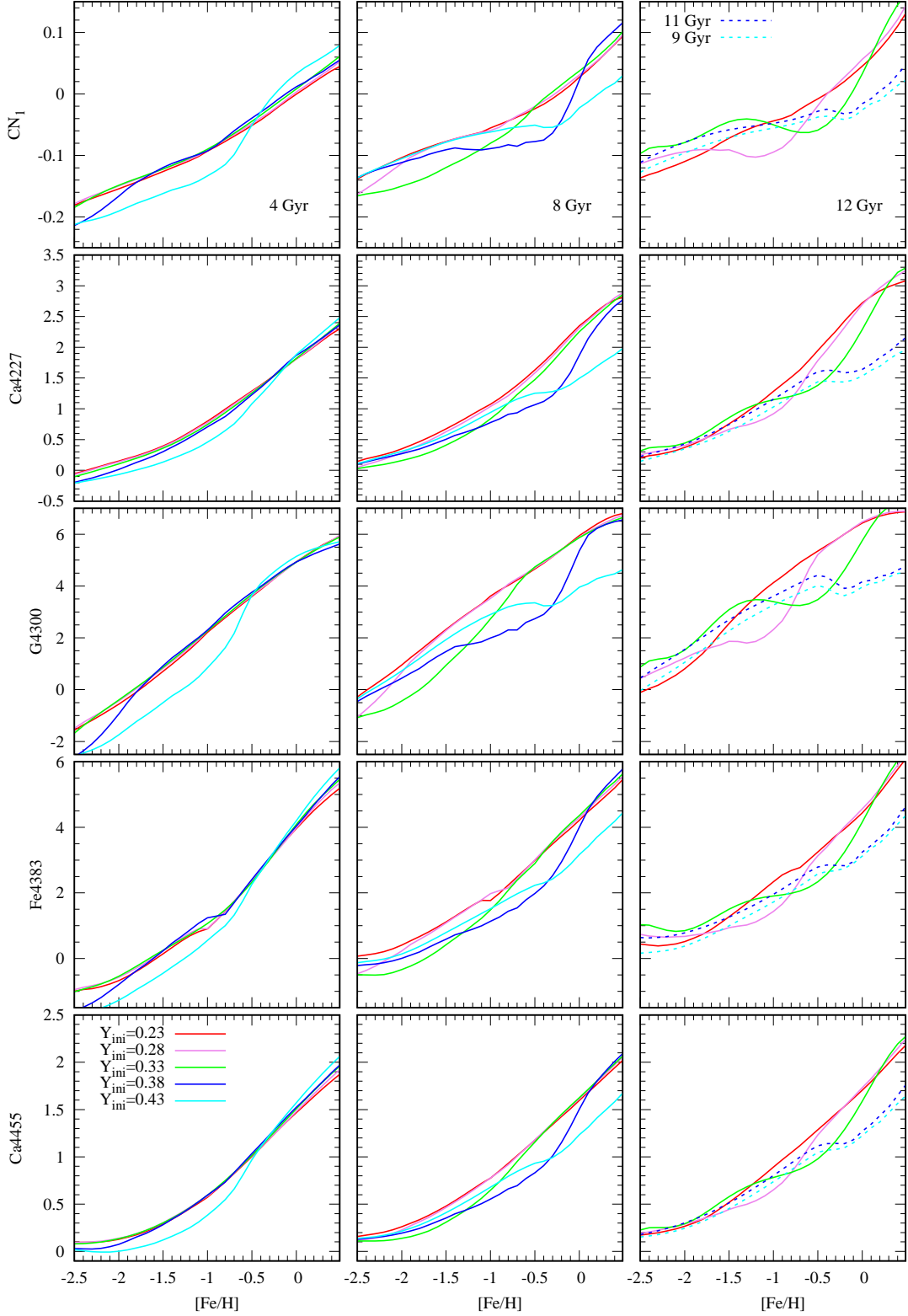}
\caption{Effect of {initial helium} on integrated absorption indices of SSPs.
The colors are the same as those for Figure~\ref{fig9}.
The ages of SSPs are 4, 8, and 12 Gyr from the left panels to the right.
SSP models for CN$_1$, Ca4227, G4300, Fe4383, and Ca4455 are presented.
{For $Y_{\rm ini}=0.38$ and 0.43 models, we present 11 and 9~Gyr models, respectively, in the panels of 12~Gyr.}
}
\label{fig11}
\end{figure*}

\begin{figure*}
\includegraphics[angle=0,scale=0.75]{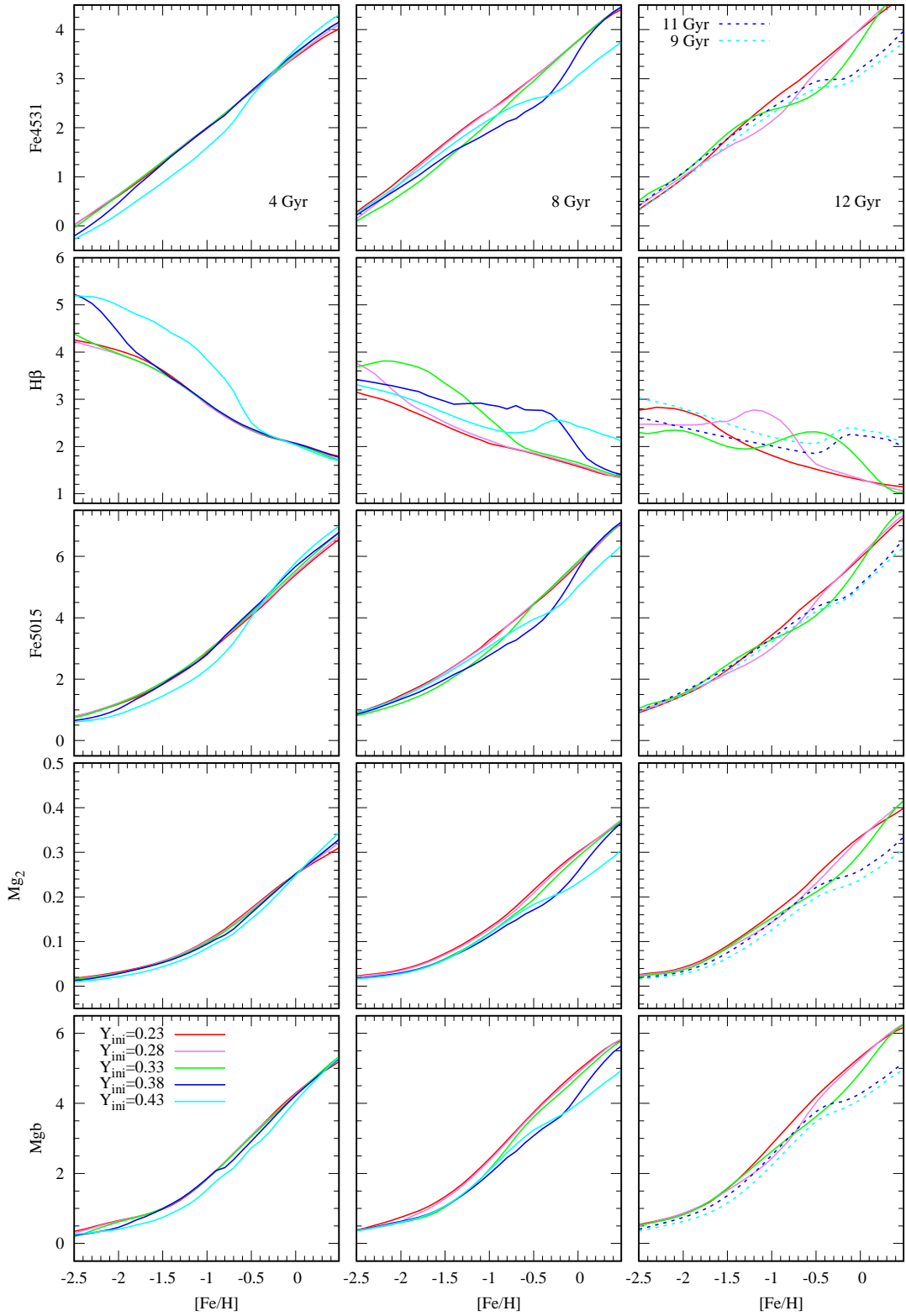}
\caption{
Same as Figure~\ref{fig11} but for indices of Fe4531, H$\beta$, Fe5015, Mg$_2$, and Mg{\it b}.
}
\label{fig12}
\end{figure*}

Recent studies \citep{Lee05, 2005ApJ...621..777P, 2007ApJ...661L..53P, Yoon08, Han09, 2010AJ....140..631B} show that the helium-enhanced populations take up $\sim$30\% of the whole population in Milky Way GCs. 
For a more quantitative comparison with observations, the helium-enhanced populations should be incorporated properly into stellar population synthesis models.
The choice of {initial helium} abundance for the helium-enhanced stellar populations is crucial because, as shown in Figures~\ref{fig9}-\ref{fig12}, the {initial helium} abundance of the helium-enhanced populations controls when the extreme HB stars come out.
Thus, the choice of the fraction of the helium-enhanced populations and the {initial helium} abundance alter the integrated colors and absorption indices of stellar populations.
This implies that, without taking into account the effect of the helium-enhanced populations, the age and metallicity measurements of remote stellar systems would substantially be either overestimated or underestimated \citep[see, for instance,][]{2005ApJ...619L.103L, 2007ApJS..173..607R, 2011ApJ...740L..45C, 2013ApJ...769L...3C}.

It is still unclear how the helium-enhanced subpopulations formed on the Milky Way GCs\footnote{The most likely processes of helium enhancement in GCs are the pollution from the intermediate-mass asymptotic giant branch stars \citep{2008MNRAS.385.2034V} and/or fast-rotating massive stars \citep{2007A&A...464.1029D}, and are due to the enrichment by supernovae \citep{2005ApJ...621..777P, 2009Natur.462..480L}.} and whether or not these populations can affect the helium enrichment in the Galactic scale.
The most plausible and effective way to link these two stellar systems together is to have star clusters with helium-enhanced stellar populations act as building blocks of their parent galaxies.
As well studied by \citet{2007ApJ...661L..49L}, the Milky Way GCs containing helium-enhanced stellar populations are generally luminous and also show peculiar kinematics compared to that of the other GCs.
This may suggest that the GCs with the helium-enhanced stellar populations have a different origin and are possible candidates for the disrupted cores of the Milky Way building blocks.
The discoveries of a large amount of sdB stars \citep{2006BaltA..15...77R} and CN-enhanced stars \citep{2010A&A...519A..14M, 2011A&A...534A.136M, 2016ApJ...825..146M, 2016ApJ...833..132F, 2017MNRAS.465..501S} in the Milky Way halo and bulge have also strengthened the view that these stars are scattered building block remnants, because the only environment where CN-enhanced stars and sdB stars can form is inside massive GCs\footnote{\citet{2017MNRAS.465..501S} discuss that N-rich stars may have been formed in similar environments of Galactic GCs but gravitationally are not bound to GCs. However, more work is needed to reveal the origin of these stars.}.
In addition, the dynamical evolution model of GCs presented in \citet{2016MNRAS.456L...1C} supports the notion that GCs in the Milky Way are building blocks of the Milky Way halo.
A recent study by \citet{2015MNRAS.453.3906L} on the double red clump found in the Milky Way bulge proposed the close connection between the Milky Way bulge and the GCs with multiple stellar populations.
If so, both early-type galaxies and bulges of disk galaxies should be affected by the presence of helium-enhanced populations, showing strong UV fluxes and enhanced Balmer indices.
Interestingly, the well-known UV-upturn phenomenon \citep[e.g.,][]{1988ApJ...328..440B, Park97, 1998ApJ...492..480Y, 1999ApJ...513..128Y, 1999ARA&A..37..603O} and enhanced Balmer indices of early-type galaxies \citep{2006ApJ...651L..93S} are in line with the prediction of helium-enhanced stellar populations as building blocks of early-type galaxies \citep{2011ApJ...740L..45C}.

Besides the helium enhancement, it is well established that the second-generation populations in the Milky Way GCs show peculiar patterns of chemical elements such as the Na--O anticorrelation and the N enhancement \citep[][and references therein]{2009A&A...505..139C, 2009A&A...505..117C, 2012A&ARv..20...50G, 2015ApJS..216...19L}.
If we assume that the second-generation stars in massive GCs (i.e., building blocks) donate a certain amount of stars to the halo of their parent galaxies via disruption, the existence of Na excess galaxies found among massive early-type galaxies can be understood \citep{2013ApJS..208....7J}.
The problem is that the effect of helium enhancement is in the opposite direction of the enhancement of the individual elements as shown in Figures \ref{fig11}-\ref{fig12}.
However, this can be understood too, given that the helium-enhanced population takes only a small fraction of the whole population and the effect of the element variation on these indices overwhelms the effect of the temperature increase.
Another intriguing aspect of the helium-enhanced stellar populations is that these populations can also influence the gravity-sensitive indices such as the Na I doublet \citep{1997ApJ...479..902S} and the Wing-Ford index \citep{1997ApJ...484..499S} due to their smaller mean mass at a given age.
If helium-enhanced populations are abundant in a galaxy or a star cluster, their smaller mass may also lead to a stronger equivalent width of the Na~I doublet  and Wing-Ford indices.
This is an important issue because the strong gravity-sensitive indices are usually interpreted as bottom-heavy stellar mass functions of early-type galaxies \citep[e.g.,][]{2010Natur.468..940V, 2012Natur.484..485C, 2012ApJ...760...71C}.
{We will deal with the issues for other element variations such as C, N, O, and Na, as well as the effect of gravity-sensitive indices on the model, in our forthcoming papers.}

\acknowledgments
{We thank the referee for helpful suggestions.
Y.W.L. and S.J.Y. acknowledge support from the National Research Foundation of Korea to the Center for Galaxy Evolution Research. 
S.J.Y. acknowledges support from Mid-career Researcher Program (No. 2012R1A2A2A01043870) through the National Research Foundation (NRF) of Korea grant funded by the Ministry of Science, ICT, and Future Planning (MSIP).
This work was supported by the National Research Foundation of Korea(NRF) grant funded by the Korea government(MSIP) (No. 2017R1A2B3002919).
This work was partially supported by the KASI-Yonsei Joint Research Program.}

\clearpage

\clearpage

\clearpage

\clearpage

\clearpage

\clearpage
\clearpage

\clearpage

\clearpage

\clearpage

\clearpage

\clearpage

\clearpage

\begin{deluxetable}{lcc}
\tabletypesize{\scriptsize}
\tablewidth{0pt}
\tablecaption{\label{tab:table1} INPUT PARAMETERS ADOPTED IN MODELS}
\tablehead{\colhead{Parameters} &\colhead{Normal-helium Model} &\colhead{Helium-enhanced Model}}
\startdata
Slope of Salpeter initial mass function, $s=x+1$ &$2.35$&$2.35$\\
$\alpha$-elements enhancement, [$\alpha$/Fe]& 0.3& 0.3\\
HB mass dispersion, ${{\sigma_{}}_{M}}$ ($M_{\odot}$)&0.015&0.015\\
Reimers' mass-loss parameter, $\eta$&0.4, 0.5, and 0.6 & 0.4, 0.5, and 0.6\\
$\eta$ calibrated for inner-halo GCs of the Milky Way & 0.59 (12~Gyr assumption)& ...\\
Initial Helium abundance, $Y_{\rm ini}$ & 0.23 & 0.28, 0.33, 0.38, and 0.43\\
Age, $t$ (Gyr) & 1 to 15 & 1 to 15\\
Metallicity coverage in [Fe/H] & $-2.5$ to 0.5 & $-2.5$ to 0.5\\
\enddata
\end{deluxetable}

\clearpage
{
\begin{deluxetable}{cccccc}
\tabletypesize{\scriptsize}
\tablewidth{0pt}
\tablecaption{\label{tab:table2} THE CONVERSION BETWEEN ${\rm [Fe/H]}$ AND $Z$ WITH RESPECT TO THE INITIAL HELIUM FOR ${\rm [\alpha/Fe]}=0.3$}
\tablehead{\colhead{${\rm [Fe/H]}$}  &\multicolumn{5}{c}{$Y_{\rm ini}$}\\
\cline{2-6}
\colhead{}  &\colhead{0.23} &\colhead{0.28} &\colhead{0.33} &\colhead{0.38} &\colhead{0.43}\\
\cline{2-6}
\colhead{}  &\multicolumn{5}{c}{$Z$}}
\startdata  
-2.5           & 0.00010        & 0.00009        & 0.00009        & 0.00008        & 0.00008        \\
-1.7           & 0.00064        & 0.00060        & 0.00056        & 0.00051        & 0.00047        \\
-0.9           & 0.00398        & 0.00372        & 0.00347        & 0.00321        & 0.00295        \\
-0.3           & 0.01515        & 0.01417        & 0.01319        & 0.01220        & 0.01122        \\
0.0            & 0.02856        & 0.02670        & 0.02485        & 0.02299        & 0.02114        \\
0.5            & 0.07279        & 0.06807        & 0.06334        & 0.05861        & 0.05389      \\ 
\enddata
\end{deluxetable}
}

\clearpage

\begin{deluxetable}{lc}
\tabletypesize{\scriptsize}
\tablewidth{0pt}
\tablecaption{\label{tab:table3} THE VALID AGE RANGE OF THE HELIUM-ENHANCED MODELS FOR $\eta=0.5$}
\tablehead{\colhead{Initial Helium ($Y_{\rm ini}$)} &\colhead{Age}}
\startdata
$0.28$ & 1 $\sim$ 15~Gyr \\
$0.33$ & 1 $\sim$ 13~Gyr \\
$0.38$ & 1 $\sim$ 11~Gyr \\
$0.43$ & 1 $\sim$ 9~Gyr \\
\enddata
\end{deluxetable}

%% This command is needed to show the entire author+affilation list when
%% the collaboration and author truncation commands are used.  It has to
%% go at the end of the manuscript.
%\allauthors

%% Include this line if you are using the \added, \replaced, \deleted
%% commands to see a summary list of all changes at the end of the article.
\listofchanges

\end{document}